\documentclass[manuscript,screen]{acmart}
 \pdfoutput=1
 
\usepackage[utf8]{inputenc}
\usepackage{subcaption}

\usepackage{multirow}
\usepackage{longtable}

\begin{document}

\title{Generating GitHub Repository Descriptions: A Comparison of Manual and Automated Approaches}

\author{Jazlyn Hellman}
\affiliation{%
  \institution{School of Computer Science, McGill University}
  \country{Canada}
  }
\email{jazlyn.hellman@mail.mcgill.ca}

\author{Eunbee Jang}
\affiliation{%
  \institution{School of Computer Science, McGill University}
  \country{Canada}
  }
\email{eunbee.jang,@mail.mcgill.ca}

\author{Christoph Treude}
\affiliation{%
  \institution{School of Computing and Information Systems, University of Melbourne}
  \country{Australia}
  }
\email{ctreude@gmail.com}

\author{Chenzhun Huang}
\affiliation{%
  \institution{School of Computer Science, McGill University}
  \country{Canada}
  }
\email{chenzhun.huang@mail.mcgill.ca}

\author{Jin L.C. Guo}
\affiliation{%
  \institution{School of Computer Science, McGill University}
  \country{Canada}
  }
\email{jguo@cs.mcgill.ca}


\renewcommand{\shortauthors}{Hellman, et al.}


\begin{abstract}
Given the vast number of repositories hosted on GitHub, project discovery and retrieval have become increasingly important for GitHub users. Repository descriptions serve as one of the first points of contact for users who are accessing a repository. However, repository owners often fail to provide a high-quality description; instead, they use vague terms, the purpose of the repository is poorly explained, or the description is omitted entirely. In this work, we examine the current practice of writing GitHub repository descriptions. Our investigation leads to the proposal of the LSP  (\textbf{L}anguage, \textbf{S}oftware technology, and \textbf{P}urpose) template to formulate \textit{good} descriptions for GitHub repositories that are clear, concise, and informative. To understand the extent to which current automated techniques can support generating repository descriptions, we compare the performance of state-of-the-art text summarization methods on this task. Finally, our user study with GitHub users reveals that automated summarization can adequately be used for default description generation for GitHub repositories, while the descriptions which follow the LSP template offer the most effective instrument for communicating with GitHub users.
\end{abstract}



\maketitle

\section{Introduction}

GitHub is one of the leading software development platforms and hosts millions of open source software repositories. It enables software developers and users with different backgrounds to `learn, share, and work together’ to `build, ship, and maintain their software'~\cite{gh_mission_2021,microsoft_gh}. In addition to managing their own projects, GitHub users often search for, utilize, and contribute to others' repositories through various functions that support social coding (e.g. stars, forking, pull requests, etc.). Regardless of users' concrete motivation and their unique information discovery process, whenever they encounter a new repository, the immediate information presented is the name of the repository and its owner, a short description about the repository, the primary programming language of the repository, the total number of stars and forks a repository has obtained, and the latest updating date. Given the large number of repositories hosted on GitHub, this information is critical for its users to decide which repositories might satisfy their needs and are worthy of further investigation. 

\begin{figure}[bpt] 
\centering \fbox{\includegraphics[scale=.30]{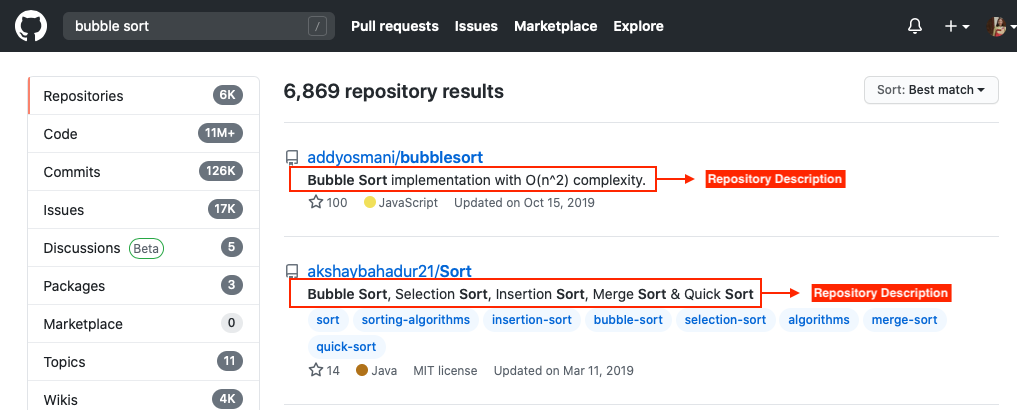}}
\fbox{\includegraphics[scale=.30]{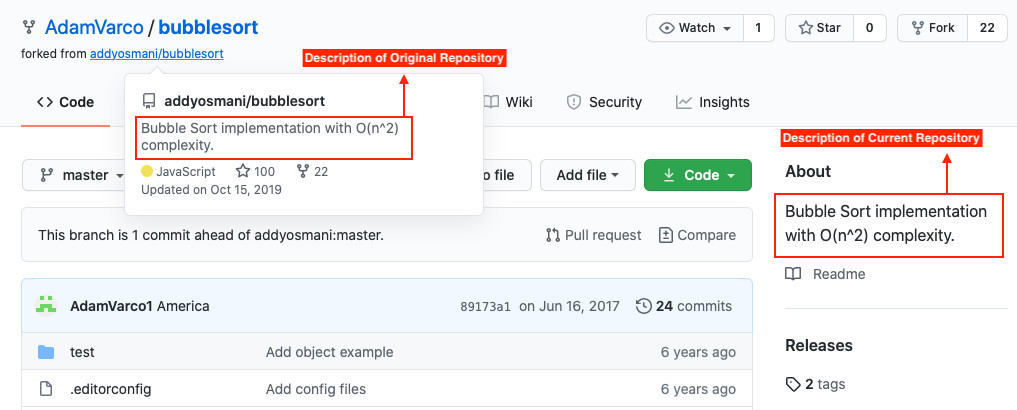}}
\caption{Example of GitHub repository descriptions rendered during repository search (above), on the repository front page (bottom right) and whenever the mouse is hovered on a link to one repository (bottom left).}
\label{fig:screenshot}
\end{figure}

Among the prompted information about a GitHub repository, the only non-automatic component is \textbf{``description''} which serves as a concise version of describing the repository. Figure~\ref{fig:screenshot} illustrates how the descriptions are rendered in different scenarios. While descriptions serve as a prominent component for a repository's documentation, the criteria for writing and maintaining descriptions are vague and lack standardization which results in many repositories’ descriptions containing outdated information, being unclear, or even empty. Due to this, users must spend considerable time further researching a repository to determine its use or they might choose to bypass a potentially useful repository altogether. 

Notable prior contributions to better facilitate information discovery on GitHub have included work on classifying contents of README files \cite{prana2019categorizing}, identifying similar repositories \cite{zhang_2017, altarawy_shahin_mohammed_meng_2018, nguyen_rocco_rubei_ruscio_2020}, and identifying non-engineered repositories from engineered repositories \cite{munaiah2017curating}. However, no prior work has focused on repository descriptions as a primary concern for project discovery. Towards this direction, our work focuses on understanding the current practices by GitHub users in writing descriptions  and investigating approaches such as guided manual creation and state-of-the-art natural language summarization techniques for creating a repository description.
Our work answers the following research questions:

\textit{\textbf{RQ1} What is the current state of practice for writing GitHub repository descriptions?}

Our empirical investigation of engineered software projects hosted on GitHub reveals that the descriptions are often not explicitly stated in their repository, and when they are there, not easy to read. While most descriptions contain some form of a repository's purpose, it is often vague, poorly communicated, or missing other important information. 
We, therefore, propose the LSP template (stands for \textbf{L}anguage, \textbf{S}oftware technology, and \textbf{P}urpose) for formulating clear, concise, and informative descriptions for GitHub repositories.


    
\textit{\textbf{RQ2} To what extent can natural language summarization techniques support automated description generation?}

We investigated a linguistic heuristic to identify if repository descriptions contain the purpose of the repository and curated a dataset using this heuristic. Using the curated dataset, we compared state-of-the-art extractive and abstractive summarization approaches for the generation of descriptions. Our experiments demonstrated that AbstractSum, which was proposed in previous work on summarizing software artifacts, yields the best performance in terms of ROUGE scores.


        
    
\textit{\textbf{RQ3} Comparing the descriptions generated or written through different approaches, what preferences do GitHub users express?}

Our user study with twelve GitHub users provides a comprehensive picture of users' preferences regarding the descriptions created through different approaches. 
The descriptions written using the LSP template are ranked highest by the users in general and when evaluated using concrete quality criteria.
The descriptions generated by the automated approaches are rated close to the descriptions provided by the repository owners in terms of content coverage and accuracy. Our study indicates the possibility of using automatically generated descriptions as a default version and suggests concrete considerations for improving the descriptions.

        


\section{Related Work}

Our work is primarily contextualized in the work of information discovery on GitHub and the application of natural language summarization techniques to software artifacts.

\textit{A. Project Discovery on GitHub.} Prior effort on repository discovery is mostly related to repository classification. Muniah et al.’s reaper \cite{munaiah2017curating} works to distinguish engineered software projects from other repositories on GitHub. An engineered repository is one that follows software engineering conventions whereas other repositories are those that do not, for example repositories containing tutorials and projects for school. Reaper utilizes the GHTorrent project which is ``an effort to create a scalable, queriable, offline mirror of data offered through the GitHub Rest API”  \cite{Gousi13}. Zhang et al.’s RepoPal \cite{zhang_2017} works to detect similar repositories using key GitHub features (such as similar README files) to help facilitate actions such as source code reuse, explore related repositories, and identify plagiarism. Altarawy et al.’s LASCAD \cite{altarawy_shahin_mohammed_meng_2018} automatically categorizes software technologies from source code and finds similar repositories based on the categorizations. Prana et al.’s multilabel classifier \cite{prana2019categorizing} processes textual features of READMEs and classifies README file contents for improved information discovery.

From the social coding perspective, previous work also investigates factors affecting repository popularity through the monitoring of related features supported by GitHub, such as stars. For example, Borges et al. \cite{7816479} identified programming language and application domain as the main factors impacting a repository's number of stars in addition to identifying four patterns of popularity growth through a time series clustering algorithm. Work such as this aims to help developers continue to grow and remain competitive. 

Comparing with existing work, our study supports project discovery from a different angle. We present a novel investigation on repository descriptions which is one of the primary information types accessed by the users during project discovery. We aim to understand what constitutes a good description to best facilitate information discovery, and what automated techniques can support the generation of such a description. 

\textit{B. Natural Language Summarization Applied to Software Artifacts.}
Natural Language Processing (NLP) has been widely adopted in software engineering research. Particularly, automatic summarization of software artifacts has shows great potential for topics such as documentation generation~\cite{wan2018improving}, information extraction~\cite{rastkar2010summarizing}, and task automation~\cite{liu2019automatic}. With respect to input and output types, summarization techniques applied to SE can also be categorized as: \emph{text-to-text, code-to-text, code-to-code, and mixed artifact summarization} \cite{moreno2017automatic}. Our work is an application of \emph{text-to-text} summarization to the creation of repository descriptions, with GitHub README contents as input and the repository description as output. 

Summarization was suggested to be used to automate and reduce manual efforts on many concrete tasks. These include efforts to generate smaller summaries of bug reports \cite{rastkar2010summarizing}, user reviews \cite{di2016would}, and pull requests \cite{liu2019automatic}. In particular, we are interested in the work by Liu et al. \cite{liu2019automatic}, where the authors propose a summarization model to automatically generate pull request descriptions from commit messages and added source code comments. Their model, while similarly motivated to improve information discovery and better facilitate expertise sharing and program comprehension, does not focus on improving the description nor identifying which specific types of information are most useful for information discovery of a repository.


\begin{figure*}[t]
\centering
\includegraphics[width=0.9\textwidth]{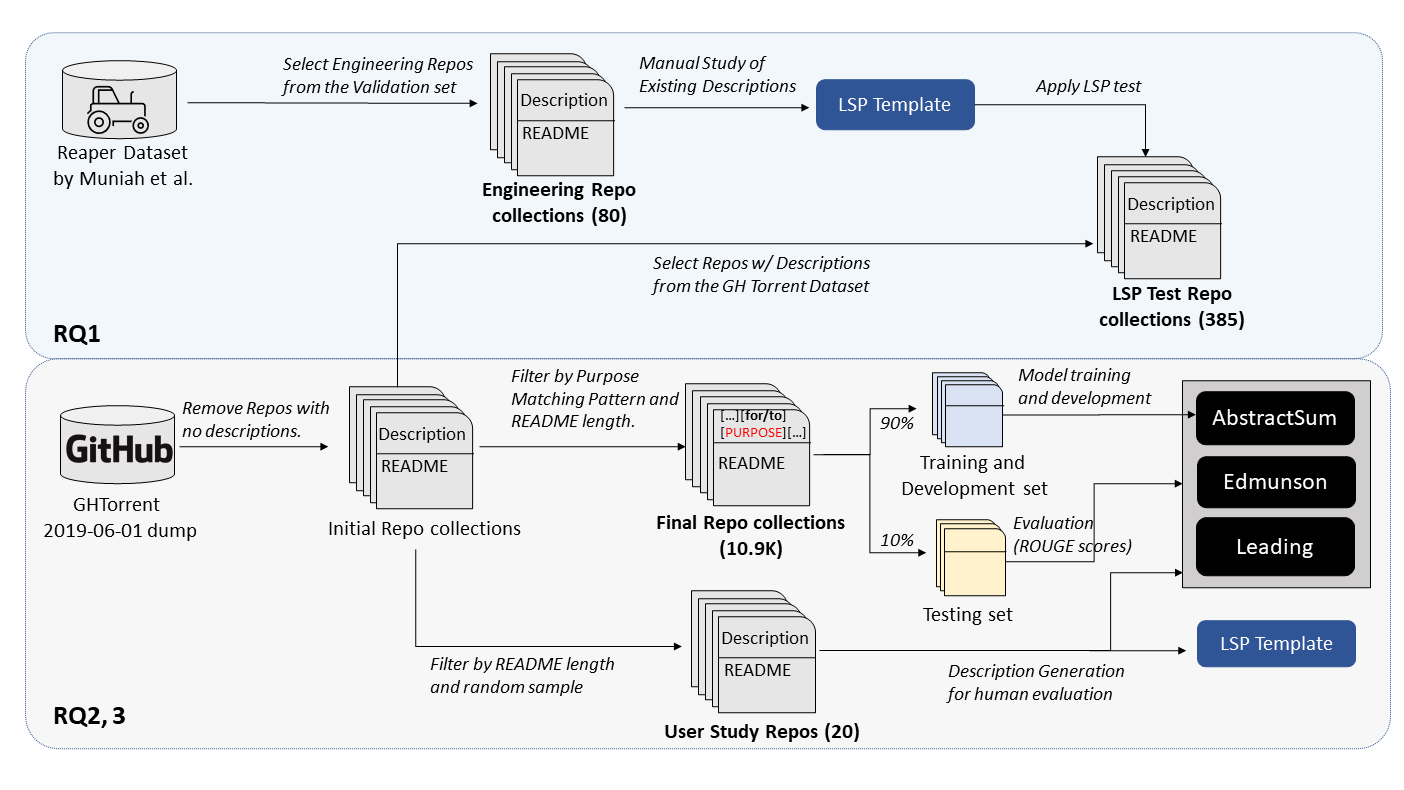}
\caption{The process of how each dataset is created and used for answering each research question. The number of repositories in those datasets is indicated in the parentheses.} 
\label{fig:data_process_overview}
\end{figure*}

\section{Exploring a 'Good' Description: A Manual Approach}
\label{sec:template}
To answer \textit{RQ1: what is the current state of practice for writing GitHub repository descriptions}, we first turned toward GitHub's own guidelines to see what precedent there was for the types of information to include in a description. According to the GitHub Guides page~\cite{GH_README_guide}, a description should be ``clear, short and to the point.''. It continues to say a description should ``describe the importance of your project, and what it does.''~\footnote{While the guide discusses the description within the README file, the suggestion is also applicable to the description we focus on (as shown in Figure~\ref{fig:screenshot}).} While we surmise this to reference the `purpose' of a repository, such a guideline is vague and subjective. Moreover, a large portion of repositories do not even provide a description at all -- that is more than 1.2M repositories out of the 3M latest active repositories from the GHTorrent 2019-06-01 dump \cite{Gousi13} utilized later in the paper (Section \ref{sec:auto}; Figure \ref{fig:data_process_overview}).

After a review of existing descriptions from samples of GitHub repositories in which we encountered a wide range in state of descriptions, we sought to understand what makes a description understandable and comprehensive to a user, and thus improves the user’s information discovery process -- specifically what the characteristics are of a ‘good’ description. Our investigation leads to the proposal of the LSP template; this section details the process of creating the template. We hypothesize this template provides a clear guide for writing descriptions that enable a user's comprehension and understanding of a repository; the template is validated in our user study detailed in Section \ref{sec:user_study}. This section discusses the template creation and an analysis of the current practice of writing GitHub descriptions through the lens of the template.

\subsection{Method}
\label{subsec:lspmethod}
We begin by creating a version of good descriptions from the perspective of the average user on GitHub through manual analysis of existing repositories and descriptions.
To select the repositories hosted on GitHub for analysis, we use the dataset curated by Muniah et al. from their work on selecting GitHub repositories with evidence of software engineering practice~\cite{munaiah2017curating}. Due to the magnitude and variance of repositories on GitHub, for this step we focused on engineered projects in order to allow for a sophisticated analysis of high quality repositories based on the assumption that they would have high quality descriptions and README files. Due to the effort required for the thorough manual analysis, we focus on a relatively small subset of 80 repositories from their validation dataset for the analysis. This base dataset was composed by a random sampling from over 1.8M active repositories at the time of Muniah et al.'s \cite{munaiah2017curating} analysis with two of their authors independently identifying engineered repositories.   

\begin{table*}[t]
    \centering
     \caption{Examples of descriptions written using the LSP template, in comparison with the original descriptions provided by the owners}
    \label{tab:LSP_examples}
    \begin{tabular}{p{0.25\linewidth} p{0.23\linewidth} p{0.45\linewidth}}
    \toprule
    \textbf{Repository owner/name} & \textbf{Initial Description} & \textbf{LSP-Description} \\
    \midrule
    AgilTec/cadenero \cite{agilTec/cadenaro} &
    Rails.API Authentication Engine for multitenant RESTful APIs &
    \color{blue} Ruby \color{black} implementation of a \color{orange} Ruby on Rails API \color{red} Authentication Engine for consumers of multitenant RESTful APIs services \\
    \hline
    brenoc/opentracks \cite{klzns/opentracks} &
    A Flask/Python client for Open-Transactions & 
    \color{blue} Python \color{black} implemented \color{orange} Flask microweb client \color{black} for the \color{red} open-source financial technology Open-Transactions project. \\
    \hline
    onaio/onadata \cite{onaio/onadata} &
    Collect, Analyze and Share &
    \color{blue} Python \color{black} implented \color{red} platform for data collection/analyzation/sharing \color{black} that utilizes the \color{orange} Sphinx and Django frameworks. \\
    \hline
    clementine-player/Android-Remote \cite{clementine-player/Android-Remote} & Control Clementine from your Android device & 
    \color{blue} Java \color{black} powered \color{orange} android application \color{black} to \color{red} remotely control \color{black} your \color{orange} open-source \color{red} music player Clementine. \\
    \hline
    rafallo/p2c \cite{rafallo/p2c} & New way to watch video! & 
    \color{blue} Python \color{black} implemented \color{orange} desktop application \color{black} for \color{red} streaming pre-downloaded content \color{black} through the \color{orange} BitTorrent communication protocol. \\
    \bottomrule
    \multicolumn{3}{@{}p{5in}}{\footnotesize \textbf{Notation:}
    \color{blue} [Language]
    \color{orange}[Software Technology]
    \color{red} [Purpose]}
    \end{tabular}
   
\end{table*}

Among the 80 repositories, 71 of them had an existing description and 73 of them contained a non-empty README file. The mean number of words per description was 9.394 words with a median of 7 words; the shortest description contained only 2 words and the longest contained 37 words. Of the 71 repositories, 7 descriptions were 2 sentences, one description was 3 sentences, and the remaining descriptions were all one sentence. Using NLTK\footnote{https://www.nltk.org/api/nltk.tokenize.html} for tokenziation, the README files had a mean of 15 sentences with a median of 9 sentences; the shortest existing README files had five words and the longest contained 3007 words.

To create a ground truth, we evaluated each repository's description and wrote new, comprehensive descriptions for every sentence. For each repository, one of our authors first referenced various information sources to build an accurate understanding of the repository, this included the README file, relevant websites linked from the README files, and the web search results when querying the repository title. Then, the same author examined the provided description and documented any edit made to improve the description, including adding or removing information, and reorganizing the sentence structure (only two repositories did not require changes to their existing description). Three of the authors then discussed those edits on individual descriptions until agreement was reached for all of them. Examples of some of these descriptions can be seen in Table \ref{tab:LSP_examples}.\footnote{\label{ft:supplement}See supplemental materials.}

\subsection{LSP Template -- Language, Software Technology and Purpose}
\label{subsec:lsp}
We observed that three primary information types were commonly included in establishing a comprehensive and understandable description: the Coding \textbf{L}anguage the repository is written in; any \textbf{S}oftware Technologies used in the repository; and the \textbf{P}urpose of the repository. Accordingly, we suggest to include those information types for writing clear, concise, and informative descriptions for the GitHub repositories, referred to as a LSP template. Table~\ref{tab:LSP_examples} presents five examples of the original description and the descriptions created using the LSP template from the target dataset.
Evidently the style of the original descriptions varies greatly. Some of them (e.g. ``Rails.API Authentication Engine for multitenant RESTful APIs'') are close to the LSP template with providing important software technologies.  However, some repositories only provide several general terms and are thus hard to interpret without context (e.g. ``Collect, Analyze and Share''). In contrast, LSP template descriptions are more readable and comprehensive (as also demonstrated in the user study detailed in Section \ref{sec:user_study}). 

Among the three primary components in the LSP template, programming language and software technology of a repository are either trivial to identify (i.e. programming language) or have been addressed in previous work (i.e. software technology~\cite{nassif2018automatically}). The purpose component, however, is so far inadequately discussed in the literature. We formally identify the \textbf{Purpose} to be the information representing `what' the user can do with a repository. From Table~\ref{tab:LSP_examples}, we can observe a clear gap between the original descriptions and the descriptions written following our LSP template due to a missing or unclear purpose in the original descriptions. 

\begin{table}[]
    \centering
    \caption{Interpretations for Fleiss' Kappa agreement \cite{landiskoch}}
    \begin{tabular}{c|c}
        Kappa Statistic & Strength of Agreement \cite{landiskoch} \\
        \hline
        $<$ 0.00 & Poor \\
        0.00-0.20 & Slight \\
        0.21-0.40 & Fair \\
        0.41-0.60 & Moderate \\
        0.61-0.81 & Substantial \\
        0.81-1.00 & Almost Perfect \\
    \end{tabular}
    \label{tab:fleissagreement}
\end{table}

\subsection{The Practice of Writing GitHub Descriptions}
\label{subsec:lsptest}
To quantitatively examine such a gap, we have developed a manual test to inspect the contents of existing descriptions called the LSP-Test. The test manually rates a repository's description across the three template categories where the rater indicates whether the description contains each category on a binary basis (0 means category is not present and 1 means category is present). To establish the test's validity, three of the authors independently rated two batches of thirty repositories randomly sampled from the 3M most recent repositories in the GHTorrent dump that contained descriptions (this was a population of \~1.8M repositories and the same initial dataset our user study was sampled from).
After the first ratings, a coding guideline was established for each category of the template. The guideline is included in the Appendices. 

\begin{figure*}[t!]
\centering\caption{LSP Test ratings, grouped by coding language of sample set, for each category of the template. Total population is 385 repositories; red represents ratings of 0 (category not present in description) and green represents ratings of 1 (category is present in description).}  
    \label{fig:LSPtest_ratings}
    \begin{subfigure}[t]{.32\linewidth}
        \centering
        \includegraphics[width=\linewidth]{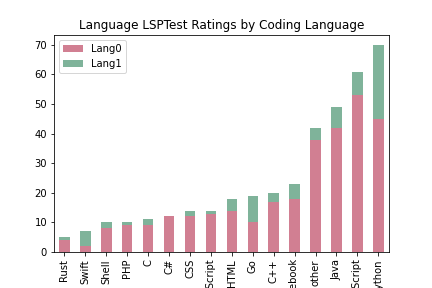}
        \subcaption{Language}
        \label{fig:lsp_lang_ratings}
    \end{subfigure}
    \begin{subfigure}[t]{.32\linewidth}
        \centering
        \includegraphics[width=\linewidth]{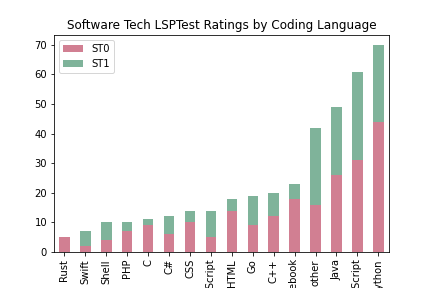}
        \subcaption{Software Technology}
        \label{fig:lsp_st_ratings}
    \end{subfigure}
    \begin{subfigure}[t]{.32\linewidth}
        \centering
        \includegraphics[width=\linewidth]{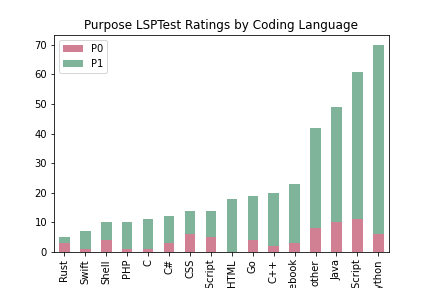}
        \subcaption{Purpose}
        \label{fig:lsp_purp_ratings}
    \end{subfigure}
    
\end{figure*}

The Fleiss' Kappa inter-rater agreement for each category is as follows: Language - 0.916; Software Technology - 0.871; and Purpose - 0.664 \cite{fleiss1971measuring}. This indicates a substantial level of agreement for the category Purpose and a near perfect level of agreement for the remaining categories \cite{landiskoch} (see Table \ref{tab:fleissagreement}). The categories of language and software technology were mostly trivial to identify; the agreement for purpose remained lower due to some edge cases that the authors all agreed were instances of subjectivity in the understanding that could not be completely eliminated. Some examples of these descriptions with disagreements were: 
    ``Add empty from edit mode"
with 2 of 3 ratings of `contains purpose'; 
    ``Swiss Army Knife for macOS"
with 2 of 3 ratings of `contains purpose'; and
    ``Mirror of Apache Subversion"
with 1 of 3 ratings of `contains purpose'. 
With the current coding guidelines for the LSP Test categories, one of the authors then coded a sample of 385 repositories randomly selected from the population of most recent ~1.8M repositories with a description from the GHTorrent dump for a 95\% confidence level and a 5\% Margin of Error.\textsuperscript{\ref{ft:supplement}} All the repository descriptions were in English; the mean number of words is 11.11 with a standard deviation of 7.94 and the mean number of sentences is 1.14 with a standard deviation of 0.43. 79 repository descriptions (20.52\%) contained a language; 167 repository descriptions (43.38\%) contained a software technology; and 317 repository descriptions (82.34\%) contained a purpose. Figure \ref{fig:LSPtest_ratings} shows the rating distributions by coding language of the repository. 

Despite the coding language being automatically specified by the GitHub software (as seen in Figure \ref{fig:screenshot} underneath the description), when it is separate from the description, this requires more memory recall on the part of the user reading descriptions and searching for specific repositories. Additionally, of the 79 repositories that had a language in their description, 8 descriptions included different or additional languages from what is listed automatically by GitHub. For example, the following description ``Thinking in PHP internals, An open book on PHP Internals" is listed for a repository with detected programming language ``HTML". A closer inspection shows that there is almost equal amounts of PHP files to HTML files in the repository; because this description is listed as an HTML repository, a user might think it is only HTML pages containing the PHP Internals book and not also code in PHP. Another example is the following description with the detected programming language listed as ``JavaScript": ``The most popular HTML, CSS, and JavaScript framework for developing responsive, mobile first projects on the web."  

As can be seen in Figure \ref{fig:lsp_purp_ratings}, a high number of repository descriptions contained the purpose (317 repositories). The final guideline for identifying these purposes was established as follows: A description contains a purpose if the sentence discusses (1) implementation of a key software concept, (2) code samples, tutorial, or a website, or (3) what the repository can be used for by an end user in layman's terms. As a result, we were inclusive in our rating process and rated descriptions as having a purpose even if that purpose was not always clear, well written and/or required a great amount of domain knowledge. This ambiguity in descriptions' purposes is reflected in the lower agreement rate between the three raters when establishing the coding guideline for this category. See Table \ref{tab:LSPTest_ratings} for some examples on descriptions and their associated ratings.

The LSP test has illustrated that of the repositories containing descriptions, only 32 repositories in our sample set (8.3\%) have descriptions containing each category of the template. Moreover, it remains that despite a high number of descriptions containing a purpose, users with insufficient domain knowledge cannot easily interpret the purpose in many descriptions, a situation often further exacerbated by poor grammar. Such observations motivate the rest of our research in considering natural language processing techniques for facilitating automatic generation of high quality descriptions and therefore support fast and clear understanding of the repository purpose.

\begin{table*}[t!]

    \centering
    \caption{Sample descriptions and ratings for the LSP test.}
    \resizebox{\textwidth}{!}{%
    \begin{tabular}{c|c|c|c|c|c}
         Name & Description  & Coding Lang & Lang & ST &  Purp \\
         \hline
         SignalR-Swift & SignalR client library written in pure Swift & Swift & 1 & 1 & 1 \\
         go-bluemix-k8s & Bluemix API with Go & Go & 1 & 1 & 0 \\
         cuberite & A lightweight, fast and extensible game server for Minecraft & C++ & 0 & 0 & 1\\
         airflow & Apache Airflow & Python & 0 & 1 & 0 \\
         balena-pihole & raspberrypi3 balenaCloud stack with Pi-hole, PADD, \& dnscrypt-proxy & Dockerfile & 0 & 1 & 0 \\
         \end{tabular}
    \label{tab:LSPTest_ratings}}
\end{table*}

\subsection{Threats and limitations}
When establishing the ground truth for good descriptions, we analyzed only 80 repositories due to the required effort to accomplish this step. These repositories were all from one dataset created in the work by Muniah et al. \cite{munaiah2017curating}. While this poses an external threat to validity, 
we argue that a deep investigation of fewer repositories was more useful to understand how to create good repository descriptions than a superficial analysis of a larger sample \cite{munaiah2017curating}. 
Another threat in this step is possible bias and human error due to only one author writing the descriptions for creating the ground truth. This threat was mitigated by two additional authors verifying the descriptions prior to the LSP template creation. Similarly, while only one author conducted the LSP test for the target population, the coding guideline used during the LSP test was verified by three of the authors with high Fleiss' Kappas for each category. 


\section{Automated Descriptions Generation}
\label{sec:auto}
During our manual inspection on repository descriptions, we observed that many descriptions contained a subset of the information contained in README files. Our exploratory investigation on overlapping terms between the descriptions and corresponding README files also raises the possibility of considering the task of automated description generation as a text summarization problem, i.e. summarizing the information from the README file. Therefore, in this section, we discuss how we approach RQ2, i.e. \textit{to what extent can natural language summarization techniques support automated description generation}. 
We start from selecting suitable natural language summarization techniques and creating a dataset of GitHub repositories that contain high quality descriptions. The performance of the automated methods is then empirically evaluated and compared using the curated dataset.

\subsection{Natural Language Summarization Techniques}
Automated text summarization techniques are actively studied by natural language processing researchers. The primary categories of such techniques include extractive and abstractive methods~\cite{mani2001automatic}. \textit{Extractive summarization} selects the most informative sentences directly from an input document based on a specified scoring mechanism. \textit{Abstractive summarization}, on the other hand, aims to summarize a body of text by generating a new string of text that conveys the most critical information. The latest abstractive methods, in particular, take advantage of recent developments in deep learning techniques and have demonstrated improved performance on several software engineering tasks such as generating code documentation~\cite{leclair2020improved} and pull request descriptions~\cite{liu2019automatic}. In this study, we explore two representative extractive algorithms and one abstractive approach for comparison.
    
\subsubsection{Extractive summarization}
\label{subsubsec:extractive_sum}
Extractive summarization techniques select one or a cluster of sentences taken exactly as they appear in the input document to output a summary. Without explicit training, some extractive methods can generate a summary through scoring sentences from the document using features such as identified topics, term distributions, and latent semantic analysis~\cite{nenkova2012survey}. Compared with the abstractive methods, they are easy to deploy and often able to achieve reasonable performance. Thus, we started by investigating more traditional extractive methods that are dominantly discussed in the NLP literature: Luhn \cite{luhn1958automatic}, Edmundson \cite{edmundson1969new}, LSA \cite{gong2001generic}, TextRank \cite{mihalcea2004textrank}, LexRank \cite{erkan2004lexrank}, SumBasic \cite{nenkova2005impact}, and KLSum \cite{haghighi2009exploring}. A preliminary study and assessment of the summarization results on the development dataset using ROUGE scores (details about the metric in Section~\ref{subsec:metrics} and dataset creation and split in Section~\ref{subsec:dataset_for_auto}) suggested the best performing method is the Edmundson algorithm. Therefore, we selected the Edmundson algorithm as the representative algorithm for extractive text summarization.

The scoring function for the Edmundson algorithm considers several factors, including term-based word frequency, user-specified cue phrases, words in the title and headings, and the location of each sentence. These features are simultaneously evaluated to weight and rank sentences to produce a summary. As a traditional extractive algorithm, it lays the foundation for many summarization techniques used today. What sets this algorithm apart from the others is that the algorithm utilizes a cue dictionary compiled using external corpora, which are statistically analyzed using linguistical concepts. The dictionary contains 139 null words (e.g., numbers, pronouns, the verb `to be'); 783 bonus words (e.g., words representing comparatives and superlatives); and 73 stigma words (e.g., anaphoric expressions such as `this, that, himself, itself'), which score sentences in different ranges of importance.

Motivated by the work on GitHub README files by Prana et al.~\cite{prana2019categorizing}, we also include a simple heuristic to generate the description of the repository, i.e. using the leading content of the README file. Their empirical study on the categories of REAMDE content reveals that the introductory and background information (the ``what'' category) normally appears at the beginning of the README, if presented. Following the same baseline setup by Liu et al. \cite{liu2019automatic}, we take the first 25 word tokens of the README and refer to this method as \textbf{Leading}. 

    

\begin{table*}[t]
   \caption{Example repositories and their descriptions from the \textit{finalRepos} dataset.}
    \label{tab:dateset_examples}

\resizebox{\textwidth}{!}{%
 \begin{tabular}{|l|l|l|}
        \bottomrule
        Matched Pattern Notation & \multicolumn{2}{l|}{{[}...{]}{[}\textbf{PTOKEN}{]}{[}{[}\textcolor{blue}{VERB}{]}\textcolor{red}{PURPOSE}{]}{[}...{]}}\\
        \hline
        PTOKEN & Purpose token indicators & ‘for’, ‘to’ \\ 
        \hline
        VERB & POS verb tags & ‘VB’,’VBD’,’VBG’,’VBN’,VBP’,VBZ’ \\ 
        \toprule
        
        \bottomrule
        \textbf{Repository owner/name} & \multicolumn{2}{l|}{\textbf{Matched Description}} \\ 
        \hline
        Mxdmedia/pyjector & \multicolumn{2}{l|}{Library \textbf{to} \textcolor{blue}{help} \textcolor{red}{control your projector} over a serial connection.} \\ 
        \hline
        dmihal/web3-provider-engine & \multicolumn{2}{l|}{A JavaScript library \textbf{for} \textcolor{blue}{composing} \textcolor{red}{Ethereum provider objects} using middleware modules.} \\ 
        \hline
        mklmzrs/react-native-forms & \multicolumn{2}{l|}{A simple, declarative API \textbf{for} \textcolor{blue}{creating} \textcolor{red}{cross-platform, native-appearing forms with React Native}} \\ 
        \toprule
    \end{tabular}
    }
\end{table*}

\subsubsection{Abstractive Summarization}
\label{subsubsec:absractive_sum}

Modern abstractive summarization is generally based on a  sequence-to-sequence framework that often requires excessive amounts of data to train. Such frameworks are comprised of an encoder that can convert the input sequence (target document for summarization) into a compact representation, and a decoder that converts the representation into the desired sequence of words as output. In some sense, abstractive summarization methods are able to synthesize the content from a document and generate a more concise summary; this process resembles how a human would perform a summarization task. The output of abstractive models is sometimes criticized for worse performance in terms of fluency and grammar and may incorporate more noise when generating inaccurate or misleading information \cite{xiao2019extractive}. 

For the abstractive summarization method, we adopt an attention based encoder-decoder model, an architecture described in Liu et al.~\cite{liu2019automatic} This work demonstrates the latest progress towards using automatic text summarization techniques for processing and generating software artifacts and the replication package with the implementation is publicly available. Liu et al. adopted a sequence-to-sequence model that summarizes a pull request in natural language from a set of commit messages and source code comments. It integrates an attention encoder decoder architecture~\cite{bahdanau2014neural} with a pointer generator~\cite{see2017_PointerGenerator} to solve the out-of-vocabulary problem, a common challenge when adopting NLP methods for software artifacts. It effectively decides whether to select word tokens from a fixed vocabulary set or to copy from the source document during the decoding step. Moreover, they adopt a reinforcement learning (RL) technique to train the model that can directly optimize the discrete evaluation metric (i.e. ROUGE metric, see Section~\ref{subsec:metrics}). The architecture is referred to as \textbf{AbstractSum} in the rest of the paper. 

We adopt the AbstractSum architecture and train the model from scratch to generate the GitHub repository descriptions by feeding the content from the README file to the model. Concretely, the encoder is a single-layer bidirectional LSTM that takes 128-dimensional word embeddings as input. The decoder is a unidriectional LSTM that shares the embedding layer with the encoder. Both encoder and decoder use 256-dimensional hidden states. We obtain a vocabulary set of 50K words from the training set specifically formulated for the experiment. Further details about the dataset preparation will be discussed in Section~\ref{subsec:dataset_for_auto}. During model training, we use the original descriptions provided by the repository owners to calculate the ROUGE-L F1 score with special loss function introduced by Liu et al. \cite{liu2019automatic}. This loss function utilizes a reinforcement learning technique called self-critical sequence training \cite{rennie2017self} and allows the ROUGE training objective to be differentiable, allowing the model to optimize for summaries that are more suitable for human evaluation.

%



\subsection{Metric Selection}
\label{subsec:metrics}
During model development and evaluation, we select from the standard ROUGE (Recall-Oriented Understudy for Gisting Evaluation) metrics \cite{lin2004rouge}. ROUGE metrics calculate the overlap between the produced summary and the reference summary (in our case, the original repository descriptions) in various ways. For example, ROUGE-N Precision is calculated as 
\[ \dfrac{total\_number\_of\_overlapping\_n\_grams}{total\_number\_of\_n\_grams\_in\_produced\_summary} \]
While the ROUGE-N Recall is calculated as  
\[ \dfrac{total\_number\_of\_overlapping\_n\_grams}{total\_number\_of\_n\_grams\_in\_reference\_summary} \]
ROUGE-L scores, on the other hand, consider the longest common sequence to avoid specifying the length of the sequences to compare. In our experiment, we use the ROUGE-L F1 score during training of AbstractSum, the same setup as Liu et al. For evaluation, to provide a more complete picture of model performance, we report F1 scores, precision, and recall for ROUGE1 (unigram), ROUGE-2 (bigram) and ROUGE-L (longest common sequence).

\subsection{Data Set Preparation}
\label{subsec:dataset_for_auto}
    
A high quality dataset is critical to the success of learning based methods. In particular, AbstractSum relies on the ``target'' descriptions to calculate the ROUGE score for optimizing the model parameters. As we discussed in Section~\ref{sec:template}, such ``target'' descriptions should contain the purpose of the corresponding repositories. This decision to focus on the purpose arose due to it being the only component of the LSP Template that generally cannot be addressed with a single term identified through keyword extraction. ` Towards this end, we adopted the following process (also summarized in Figure~\ref{fig:data_process_overview}) to select the repositories hosted on GitHub and create the training and testing datasets. Our final dataset totals 10.9K repositories with these target descriptions.\textsuperscript{\ref{ft:supplement}}

\begin{table*}[hbt]
 \centering
\caption{Summarization results against original description }\label{tab:sum_result}
\begin{tabular}{lrrr|rrr|rrr}\toprule
&\multicolumn{3}{l}{ROUGE-1} &\multicolumn{3}{l}{ROUGE-2} &\multicolumn{3}{l}{ROUGE-L} \\\cmidrule{2-10}
&F1 Score &Precision &Recall &F1 Score &Precision &Recall &F1 Score &Precision &Recall \\\midrule
Leading &40.10\% &58.75\% &33.18\% &26.89\% &40.08\% &22.30\% &36.37\% &53.36\% &30.08\% \\
Edmundson &41.99\% &58.40\% &38.65\% &28.68\% &39.88\% &26.41\% &38.59\% &53.30\% &35.59\% \\
AbstractSum &48.26\% &55.57\% &47.01\% &32.41\% &37.47\% &31.30\% &45.63\% &52.51\% &44.35\%  \\
\bottomrule
\end{tabular}
\end{table*}

We first selected the latest active one million repositories from the GHTorrent \cite{Gousi13} mysql-2019-06-01 dump and removed all the repositories that do not contain a description as our initial dataset. To determine if the descriptions might include the information about repository purposes, we then applied a syntactic pattern, which was identified when creating the LSP ground truth descriptions, on the description which is processed by a Part of Speech (POS) tagger.
The \textbf{purpose matching pattern} is defined as:
\begin{multline}
    (w_i == ``for/to'') \land (POS(w_{i+1}) \in VERB) \\
        \land ((index(<EOS>) - i) >= min\_length), \\  1<i<index(<EOS>) 
\end{multline}
In the pattern, $w_i$ is the $i$th word in that sentence, $<EOS>$ represents the end of sentence token, $VERB$ is the set that contains all the POS tags indicating verb forms, and $min\_length$ is a pre-determined parameter and set as two in our experiment. We use the $VERB$ set rather than limiting to only certain grammatically correct verb forms to accommodate potential grammatical errors in the description that happen frequently for GitHub repositories. If this pattern is matched, we consider the purpose of the repository is included in the verb phrase after the term ``for'' or ``to''. Examples of applying the purpose matching pattern are shown in Table \ref{tab:dateset_examples}.

While complex syntactic parsers are more likely to provide reliable analysis on the sentence structure, considering the nature of our dataset (i.e. sentences are often incomplete or grammatically incorrect), our heuristic pattern is sufficiently accurate and fast. We validate the effectiveness of this pattern by manually annotating a random subset of 30 repositories from the dataset by three of the authors independently. Each repository was rated on whether the purpose is conveyed. We achieved free-marginal multi-rater kappa of 0.82 and 91.11\% overall agreement, illustrating an `excellent' agreement~\cite{randolph2005free}. The manual annotation suggests a 96.7\% precision when using our purpose matching pattern to filter repository descriptions.

\begin{figure}[t!]
\centering
    \begin{subfigure}[t]{.44\textwidth}
        \centering
        \includegraphics[width=0.95\textwidth]{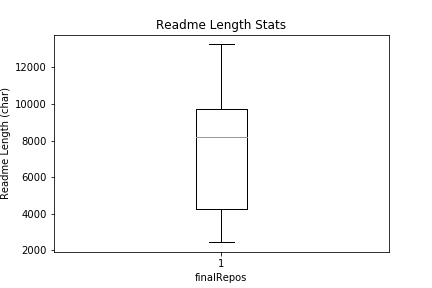}
        \subcaption{README}
        \label{fig:README_len_dataset}
    \end{subfigure}
    \begin{subfigure}[t]{.44\textwidth}
        \centering
        \includegraphics[width=0.95\textwidth]{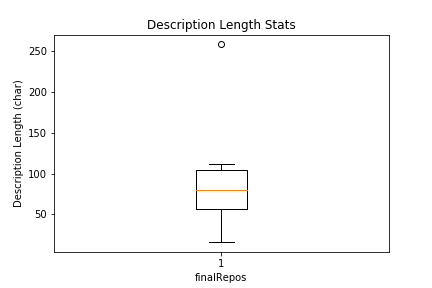}
        \subcaption{Description}
        \label{fig:desc_len_dataset}
    \end{subfigure}
    \caption{Distribution of character length for the Descriptions and READMEs in \textit{finalRepos} dataset.}  
    \label{fig:dataset_details}
\end{figure}

With establishing the high reliability of the purpose matching pattern, we next applied this pattern on each sentence in the descriptions provided by each repository and removed ones that failed to match the pattern. Considering the requirement on the input document for automated summarization methods, we further removed the repositories without README files; from the remaining repositories, we selected the subset with long README files. Based on preliminary exploration, we set the threshold for minimum README length to be in the 62nd percentile (i.e., keeping the 3/8 longest READMEs). 
This process yields a dataset with 10.9K repositories, each providing a README-Description pair, referred to as \textit{finalRepos} dataset. For training the AbstractSum method, we use 80\% of randomly sampled repositories from \textit{finalRepos} dataset, 10\% as a development set to select extractive summarization methods and tune the hyper parameters for AbstractSum.  The remaining 10\% of the data is used evaluate and compare the three automated methods. The distribution of length for the descriptions and READMEs in \textit{finalRepos} dataset is shown in  Figure~\ref{fig:dataset_details}.

\subsection{Experiment Results}

For comparing the performance of different methods, we calculated the ROUGE scores against the original repository description written by the repository owner (summarized in Table \ref{tab:sum_result}). The repeated measures ANOVA with post-hoc Tukey HSD test results on each ROUGE score show that the confidence intervals between the three methods do not overlap, indicating a significant difference among the performance of the three methods, i.e. Leading, Edmundson, and AbstractSum. Among them, the extractive summarization methods (Edmundson and Leading) tend to perform better in terms of Precision metrics, while the AbstractSum obtains largely better ROUGE scores for Recall. Overall, AbstractSum achieves stronger performance, obtaining the best F1 scores of 48.26\%, 32.41\%, and 45.63\% for ROUGE-1, ROUGE-2, and ROUGE-L respectively.




\subsection{Threats and Limitations}
The process of applying the Purpose Matching pattern might introduce noise in the data given that it is impossible to reach perfect precision. At the same time, we might miss descriptions that convey the purpose without satisfying our pattern. The high precision of the pattern (96.7\%) and the sufficient number of repositories after filtering (10.9k) suggest that our method for dataset creation offers an adequate balance. Furthermore, when validating our pattern identification outlined in Section \ref{subsec:dataset_for_auto}, a small portion of descriptions sampled were marked as vague even though they conveyed the purpose. This was often due to assumed domain knowledge or poor grammar. Therefore, the performance of AbstractSum might be improved in the future if there are more reliable ways to create the training data with enhanced quality. For the evaluation, the ROUGE score comes with challenges even though it is the standard metric for summarization. Since ROUGE measures the number of overlapping text units, it might score lowly when summaries are paraphrased despite successfully capturing salient information from the source.

\section{User Preference on Repository Descriptions}
\label{sec:user_study}
Automated evaluation metrics such as ROUGE scores are effective to compare the performance of different methods on a large scale, but fall short on providing insights on how GitHub users perceive the descriptions provided through different means. In this section, we present our design and results of a user study to address \textit{RQ3: comparing the descriptions generated or written through different approaches, what preferences do GitHub users express.}  Additionally, we discuss the validity of the LSP template from the user study results.

 \begin{figure}
     \centering
      \fbox{\includegraphics[width=0.45\textwidth]{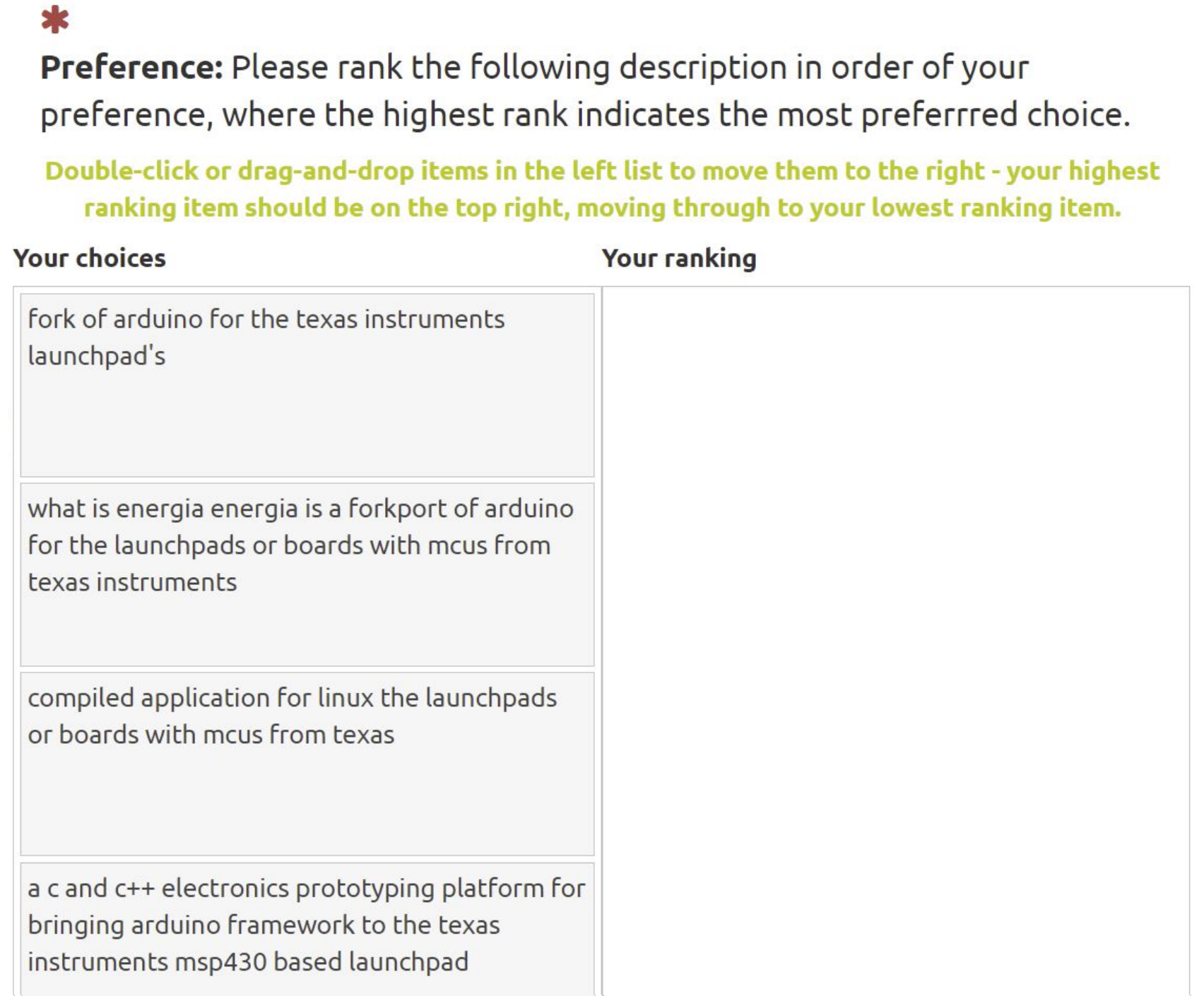}}
     \caption{Example of the user study interface for ranking the descriptions based on overall preference.}
     \label{fig:userstudy}
 \end{figure}

\subsection{Study Design}
\label{subsec:userstudy_design}
In this study, we recruit 12 GitHub users to review and rank different versions of descriptions for GitHub repositories, i.e. the original descriptions written by the repository owners (ORI),  the descriptions written by the authors of this work following the LSP template (LSP),  extractive summarizer Edmundson (EDM), and the abstractive summarizer AbstractSum (ABS). To select the repositories to review, we randomly sample twenty repositories from the initial repository dataset after removing the repositories without descriptions and those for which the length of their README files did not reach the threshold introduced in Section~\ref{subsec:dataset_for_auto}.\textsuperscript{\ref{ft:supplement}} We deliberately choose not to apply the Purpose Matching Pattern in this case since the resulting repository population is closer to what GitHub users would encounter. To set up the repositories for the user study, we forked the sampled repositories, reverted to the latest commit as of the baseline dump used in Section \ref{subsec:dataset_for_auto} and remove the description entirely.

During the study, the participants first complete a pre-survey about their demographic information, their proficiency in English, and their GitHub experience. In particular, we ask the participants about the frequency and purpose of their access to GitHub, their expected information to be included in repository descriptions, and any challenges they have experienced when reviewing descriptions. For the main task of the study, we ask each participant to evaluate the descriptions for five repositories one by one so that every repository in the study would receive three unique assessments. Following the completion of the main survey, participants complete a post-survey about their thoughts and experiences in completing the main survey.

For each repository, we ask the participants first to write the descriptions using their own words (without referencing the original description) before evaluating the provided descriptions. This task was set for two reasons, to ensure the participant conducts a thorough inspection of the repository prior to the later assessment on the provided descriptions and to inspect what information a participant will include themselves. The participant can use any online resources they presume to be relevant when writing the description. Following this task, the participant then reviews and ranks the four provided versions of descriptions based on four different dimensions to provide a comprehensive assessment, including \textbf{overall preference}, the \textbf{coverage} of each description in describing the repository, the \textbf{accuracy} of each description to the contents of the repository, and each description's \textbf{readability}~\cite{ganesan2010opinosis, graham2015re}.  The participant is also asked to provide the reason for their preference ranking, and any resources they utilized during the assessment. During the ranking, the participants have no knowledge of which approaches were used to generate each version of the descriptions. As an example, Figure \ref{fig:userstudy} is a screenshot of the user study for ranking different descriptions based on preference. A detailed presentation of the user study instructions, studied repositories,  survey questions, and associated answer options (when relevant) is available in the supplemental material.\textsuperscript{\ref{ft:supplement}}


\subsection{Study Result}
\subsubsection{Pre-Survey Results} We recruited a total of twelve GitHub users from both social media and personal contacts, all of whom had to have English proficiency. Of the 12 participants, only 3 participants indicated their primary language was not English. The participants included undergraduate students (4), graduate students (6), and industry professionals (2), each with considerable amount of GitHub experience. Concretely, 7 participants use GitHub on a daily to weekly basis and an additional 2 participants use GitHub at least a few times a month. The remaining two participants reported use of GitHub a few times a year. Moreover, 9 participants have read at least 5 repository descriptions in the last three months, and half of them have experience writing descriptions for their own repositories. The self-reported background in software development amongst the participants ranged from at least two years of university level experience to many years working in industry. All participants reported experience with software development in practice (e.g., research, contributing to open source software, web applications, video games, etc.).

All participants expressed challenges when reading repository descriptions in the past due to incorrect and unclear descriptions as well as outdated information present in the description. Moreover, they described that absence of descriptions also makes it hard to understand the purpose of the repository in consideration.
    
From the provided options (see the supplemental material for more details), the participants responded that the type of information they expect to find in a GitHub repository description includes (in an order of descending frequency): 
    \begin{itemize}
    \item Purpose of repository
    \item Software technology/ies utilized in repository
    \item Important Links (e.g. official website)
    \item Programming language of repository
    \item Maintenance development status of repository (i.e. Inactive, Dormant, etc.)
    \end{itemize}
 
Among those information types, ``purpose of repository'' is selected by all participants, indicating that it is considered as the primary content in the descriptions by the GitHub users. Among the above options, ``important links'' is the only type that is not covered in either our LSP template nor in the preview information selected by GitHub (see Figure~\ref{fig:screenshot}). While external links are not always present, they are considered important for applicable repositories.

\begin{figure}[t]
    \centering
    \begin{subfigure}[b]{0.43\textwidth}
        \centering
        \includegraphics[width=\textwidth]{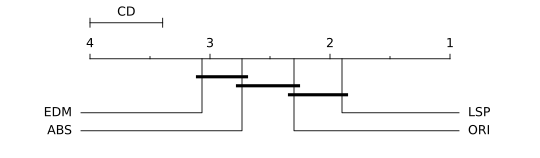}
        \caption{Preference}
        \label{fig:cd_pref}
    \end{subfigure}
    \hfill
    \begin{subfigure}[b]{0.43\textwidth}
        \centering
        \includegraphics[width=\textwidth]{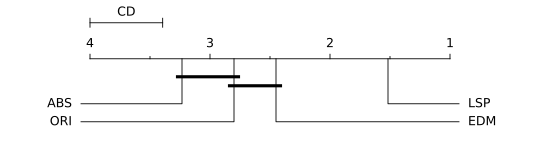}
        \caption{Coverage}
        \label{fig:cd_cov}
    \end{subfigure}
    \hfill
    \begin{subfigure}[b]{0.43\textwidth}
        \centering
        \includegraphics[width=\textwidth]{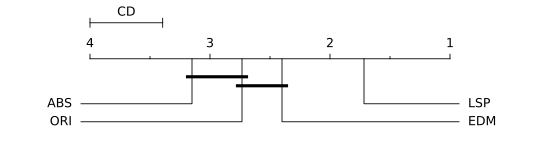}
        \caption{Accuracy}
        \label{fig:cd_acc}
    \end{subfigure}
    \hfill
    \begin{subfigure}[b]{0.43\textwidth}
        \centering
        \includegraphics[width=\textwidth]{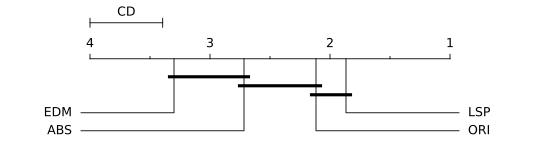}
        \caption{Readability}
        \label{fig:cd_read}
    \end{subfigure}
    \caption{Critical Distance diagrams for the post-hoc Nemenyi test. For any pair of two methods, they are considered significantly different if the difference between their average ranks is bigger than the critical distance (the length of CD).}
    \label{fig:critical_distance}
\end{figure}

\subsubsection{Main-Survey Results} We obtained a total of sixty rankings (twenty repositories, the descriptions of each ranked by three participants) along with the explanations of the rankings in free text format. On this set of rankings, we utilize the Autorank statistics package~\cite{Herbold2020} to first check if the rankings are normally distributed and then perform the appropriate Friedman test aiming to detect any differences across different descriptions~\cite{page1963ordered}. Once the differences among the groups are observed, we applied the Nemenyi post-hoc test for conducting pair-wise comparison~\cite{pereira2015overview, demvsar2006statistical}.  
    
    
Across all four dimensions for ranking (preference, coverage, accuracy, and readability), the null hypothesis that each population is normal is rejected. 
The non-parametric Friedman test as omnibus test is used to determine if there are any significant differences between the median values of the populations. The post-hoc Nemenyi test is then used to infer which differences are significant and the result is presented as critical distance diagrams in Figure~\ref{fig:critical_distance}. In those diagrams, the horizontal axis represents the value of mean rank. If there is a line connecting different approaches, it indicates that we cannot reject the hypothesis that their average ranks are equal.

    
        

         
    
        
    
Friedman test results demonstrate a significant difference in users' preference across different versions of the descriptions. Across all four dimensions, the manually written description following our LSP template is constantly ranked the highest. In particular, on the dimension of \textit{coverage} and \textit{accuracy}, the manual LSP descriptions are significantly ranked higher than all other versions, including the original description provided by the repository owners. For the automated techniques, descriptions generated by Edmundson are significantly ranked higher than AbstractSum for \textit{coverage} and \textit{accuracy}, but lower than AbstractSum on the dimensions of \textit{readability} and the general \textit{preference}.

\subsection{Observations}
\label{subsec:obs}
In addition to quantitatively measuring the users' preferences across the four dimensions, we further analyze their free text input about their reasoning process, in particular, how their criteria overlap with the dimensions we use and the resources they used to make the assessment. 

For the resources used to evaluate the description, in 11 out of 12 responses, the participants mentioned README files as the primary resource (this question was asked only once at the end of the study). Four of them suggested that links provided in the README files are also relevant and useful. Other resources mentioned  include the Wiki pages hosted on GitHub, files from the repositories close to the root directory, and Google search result on unfamiliar terms. 

When explaining their general preference, all twelve participants mentioned that descriptions should be \textbf{`short'} or \textbf{`concise'} in 19 total responses. Note that this question was posed to each participant 5 times after they ranked a repository's descriptions on the criteria of preference. 
    Participant 11 said about repository ID 11: 
    \begin{itemize}
        \item []\textit{I prefer short and concise descriptions that briefly explain the repository. Longer and detailed descriptions are helpful but I'd rather read the README for richer information.}
    \end{itemize} 
    
25 responses from ten participants mentioned the reason for their general preferences had something to do with some degree of the \textbf{readability}, including clarity, grammar, or understandability of the descriptions, and \textbf{coverage}. This explanation aligns with our quality dimensions.
    Participant 9 said about repository ID 15: 
    \begin{itemize}
        \item[] \textit{My first choice gives the idea clearly. My second choice gives all the finer details but still isn't great on the grammar and has too much unneeded information. My third choice is generated poorly and while it has the right idea, misses a lot of the key words. My last choice misses the point completely.}
    \end{itemize} 
    Participant 1 said about repository ID 5: 
    \begin{itemize}
        \item[] \textit{First one is clear. Second one gives important details but misses JavaScript. Third one barely gives any details. Last one doesn't make sense.}
    \end{itemize} 

While overall a shorter description is preferred, several participants mentioned that uncommon \textbf{acronyms} should be explained in the description. For example, participant 4 described the descriptions for repository ID 19 as follows:
    \begin{itemize}
        \item[] \textit{[...] short and to the point, didn't spell out ASR, I didn't know that acronym [...]}.
    \end{itemize}
    Participant 12 further explained the problem with searching on GitHub when using acronyms for repository ID 19:
    \begin{itemize}
        \item[] \textit{is also accurate but doesn't explain what ASR is, if i'm searching for "automatic speech recognition" I might miss it if I didn't recognize the acronym.}
    \end{itemize}
    
Regarding the \textbf{accuracy} of the descriptions, the participants raised two concerns when they responded to the overall problems they have observed during the study, including selecting irrelevant information in the description and incorrectly synthesizing information. The answer from Participant 3 exemplifies those concerns:
    \begin{itemize}
        \item[] \textit{Some of the descriptions were simply off the mark, focusing on details that were not relevant for a brief description. In other cases, it seemed like different bits of correct information were put together in misleading ways. [...] }
    \end{itemize}

Participant 1 also raised this importance of accuracy in comparison with coverage.
    \begin{itemize}
    \item[] \textit{ [...] Sometimes it felt okay if the information was not sufficient, but pointing in the wrong direction is not useful at all. [...]}\end{itemize}
 
Finally, the importance of providing high-quality \textbf{documentation} for the repository (including both the README file and description) is emphasized by all  participants when responding to the question of what difficulties they encounter in identifying the most suitable GitHub repositories in their own work. For example, Participant 3 said:
  \begin{itemize}
    \item[] \textit{The biggest problem I face is lackluster documentation, specifically on how to work with the library. Lack of guidance or outdated guidance on how to work with a library is a big time sink for me as it requires digging into the source code more deeply than usual to figure out how things work.}
    \end{itemize} 

Participants 7 suggested that documentation is related to \textbf{trust} since the quality of the documentation is considered as a proxy for the quality of the repository:
\begin{itemize}
    \item[] \textit{In general, I believe that if the README/description is not clear and well written, I don't trust the work inside. [...]}
    \end{itemize} 

Overall, our user study illustrates the importance of providing concise, accurate and easy-to-understand descriptions for GitHub repositories. The results from this study indicate the validity of the LSP template. Descriptions following the LSP template are preferred to the automated methods for the aforementioned qualities. In addition, the LSP template is preferred to the descriptions provided by the owners on the dimensions of coverage and accuracy. When selecting automated methods, a trade-off between readability and accuracy might need to be balanced. This will be further discussed in Section \ref{sec:discussion}.

\subsection{Threats and Limitations}
During the user study, we asked participants to rank different descriptions without the knowledge of the approaches used to generate each description. However, descriptions were presented in the same order for each repository. This might have led to biased responses in the rankings. 
Another threat to validity is that participants could have returned to the Main Survey part 1 where they were asked to write their own descriptions after reading the proposed descriptions or they could have searched for the original repository to get the original description for part one. Both of these actions would potentially bias the participant's written descriptions and the rankings for the remaining questions. The last threat to validity is the representativeness of the user study task. Since our participants were not actually looking for repositories with particular characteristics at the time of our study, but only did so because we asked them to, results might differ if we had been able to observe participants interact with the descriptions during their regular work.

\section{Discussion and Conclusion}
\label{sec:discussion}

    
    \subsection{Recommendations}
    As was identified in Section \ref{subsec:lsptest}, the current state of description writing on GitHub has led to a wide variety in description quality. In our investigation of manual and automatic methods for improving the overall description quality of repositories on GitHub, a number of insights were gleaned. First, our research results lead us to strongly encourage GitHub repository maintainers and contributors to keep clear and comprehensive documentation, including both the description and the README file. Not only will this inspire better trust and understanding for the users through transparency and active maintenance, but up-to-date README files will also have the potential for repository owners to leverage automatic summarization methods (see the third recommendation). 
    
    Second, while we encourage individuals to utilize the LSP template, especially because of its high rankings across the four dimensions, if this is not possible, we insist repository owners at least include the \textbf{purpose} of the repository in the description. For clear and readable purposes, we suggest using the linguistic pattern identified in Section \ref{subsec:dataset_for_auto}.
    In compliance with our observations in Section \ref{subsec:obs}, we emphasize the importance of correct grammar for maximizing readability and encourage repository contributors to be aware of the terminology they are using in their description and how this may affect readability. If a repository requires domain-specific knowledge, we recommend contributors to avoid acronyms and vague or obscure terminology without further context in the description sentence.
    
    Third, due to the lack of significant difference on the \textit{preference} and \textit{readability} dimensions between the original descriptions and the AbstractSum descriptions, we recommend the application of the AbstractSum summarization method to repositories with README files as a baseline suggestion to minimize the efforts of repository owners to keep up-to-date descriptions. However, it is important to mention that the LSP descriptions were significantly ranked higher across all dimensions and thus we defer back to recommendation two.
    
    Lastly, we recommend that descriptions remain concise and accurate, and do not include information that is not relevant to achieving the purpose of the repository. Due to the aforementioned trade-off between \textit{preference/readability} and \textit{coverage/accuracy}, until further research can be conducted, we refer to a key observation from Section \ref{subsec:obs} and recommend users to prioritize short, readable descriptions with more information available in the README file.
    
    
    \subsection{Future Work}
    Our future work includes further investigation of the trade-offs identified between readability and accuracy of descriptions, specifically in our automated methods in order to maximize both dimensions. We will also revisit and iterate the LSP template with the insights gained from the user study results and observations. Lastly, we plan to implement a user-friendly way to utilize an automatic description generator based on the outlined recommendations.
    
    \subsection{Conclusion}
    In an effort to improve information discovery on GitHub, we perform a novel investigation on GitHub repository descriptions. First, we investigate the current state of descriptions on GitHub. Our analysis concluded that the current state of descriptions is poor due to the lack of repositories' purposes included in the existing descriptions. To guide the formulation of clear, concise, and informative descriptions, we proposed the LSP template for writing their descriptions. Second, we investigated the extent to which natural language summarization techniques can support automated description generation through the creation of a novel dataset using a linguistic heuristic for identifying descriptions which contain repositories' purposes.  Finally, we contribute a user study with twelve GitHub users to obtain a comprehensive understanding of preferences on four different description approaches (the original description, the LSP description, the AbstractSum description, and the Edmundson description). Our user study verified the need for improved information discovery by way of descriptions. We conclude that in terms of overall preference and readability, the LSP template is ranked the highest. However, in terms of content coverage and accuracy, there is no critical difference between the original descriptions and the automated descriptions according to the study participants. This indicates the possibility for utilizing automated techniques in place of manual written descriptions for improving information discovery on GitHub.
    
\bibliographystyle{plain}
\bibliography{reference}

\begin{thebibliography}{10}

\bibitem{agilTec/cadenaro}
\url{https://github.com/AgilTec/cadenero}.

\bibitem{klzns/opentracks}
\url {https://github.com/klzns/opentracks}.

\bibitem{onaio/onadata}
\url{https://github.com/onaio/onadata}.

\bibitem{clementine-player/Android-Remote}
\url{https://github.com/clementine-player/Android-Remote}.

\bibitem{rafallo/p2c}
\url{https://github.com/rafallo/p2c}.

\bibitem{altarawy_shahin_mohammed_meng_2018}
Doaa Altarawy, Hossameldin Shahin, Ayat Mohammed, and Na~Meng.
\newblock Lascad : Language-agnostic software categorization and similar
  application detection.
\newblock {\em Journal of Systems and Software}, 142:21–34, 2018.

\bibitem{bahdanau2014neural}
Dzmitry Bahdanau, Kyunghyun Cho, and Yoshua Bengio.
\newblock Neural machine translation by jointly learning to align and
  translate.
\newblock {\em arXiv preprint arXiv:1409.0473}, 2014.

\bibitem{7816479}
H.~{Borges}, A.~{Hora}, and M.~T. {Valente}.
\newblock Understanding the factors that impact the popularity of github
  repositories.
\newblock In {\em 2016 IEEE International Conference on Software Maintenance
  and Evolution (ICSME)}, pages 334--344, 2016.

\bibitem{demvsar2006statistical}
Janez Dem{\v{s}}ar.
\newblock Statistical comparisons of classifiers over multiple data sets.
\newblock {\em The Journal of Machine Learning Research}, 7:1--30, 2006.

\bibitem{di2016would}
Andrea Di~Sorbo, Sebastiano Panichella, Carol~V Alexandru, Junji Shimagaki,
  Corrado~A Visaggio, Gerardo Canfora, and Harald~C Gall.
\newblock What would users change in my app? summarizing app reviews for
  recommending software changes.
\newblock In {\em Proceedings of the 2016 24th ACM SIGSOFT International
  Symposium on Foundations of Software Engineering}, pages 499--510, 2016.

\bibitem{edmundson1969new}
Harold~P Edmundson.
\newblock New methods in automatic extracting.
\newblock {\em Journal of the ACM (JACM)}, 16(2):264--285, 1969.

\bibitem{erkan2004lexrank}
G{\"u}nes Erkan and Dragomir~R Radev.
\newblock Lexrank: Graph-based lexical centrality as salience in text
  summarization.
\newblock {\em Journal of artificial intelligence research}, 22:457--479, 2004.

\bibitem{fleiss1971measuring}
Joseph~L Fleiss.
\newblock Measuring nominal scale agreement among many raters.
\newblock {\em Psychological bulletin}, 76(5):378, 1971.

\bibitem{ganesan2010opinosis}
Kavita Ganesan, ChengXiang Zhai, and Jiawei Han.
\newblock Opinosis: A graph based approach to abstractive summarization of
  highly redundant opinions.
\newblock In {\em Proceedings of the 23rd International Conference on
  Computational Linguistics (Coling 2010)}, pages 340--348, 2010.

\bibitem{gh_mission_2021}
GitHub.
\newblock About.
\newblock \url{https://github.com/about}.

\bibitem{GH_README_guide}
GitHub.
\newblock {Documenting your projects on GitHub}.
\newblock \url{https://guides.github.com/features/wikis/}, 2016.
\newblock [Online; accessed 16-March-2021].

\bibitem{gong2001generic}
Yihong Gong and Xin Liu.
\newblock Generic text summarization using relevance measure and latent
  semantic analysis.
\newblock In {\em Proceedings of the 24th annual international ACM SIGIR
  conference on Research and development in information retrieval}, pages
  19--25, 2001.

\bibitem{Gousi13}
Georgios Gousios.
\newblock The ghtorrent dataset and tool suite.
\newblock In {\em Proceedings of the 10th Working Conference on Mining Software
  Repositories}, MSR '13, pages 233--236, Piscataway, NJ, USA, 2013. IEEE
  Press.

\bibitem{graham2015re}
Yvette Graham.
\newblock Re-evaluating automatic summarization with bleu and 192 shades of
  rouge.
\newblock In {\em Proceedings of the 2015 conference on empirical methods in
  natural language processing}, pages 128--137, 2015.

\bibitem{haghighi2009exploring}
Aria Haghighi and Lucy Vanderwende.
\newblock Exploring content models for multi-document summarization.
\newblock In {\em Proceedings of Human Language Technologies: The 2009 Annual
  Conference of the North American Chapter of the Association for Computational
  Linguistics}, pages 362--370, 2009.

\bibitem{Herbold2020}
Steffen Herbold.
\newblock Autorank: A python package for automated ranking of classifiers.
\newblock {\em Journal of Open Source Software}, 5(48):2173, 2020.

\bibitem{landiskoch}
J.~Richard Landis and Gary~G. Koch.
\newblock The measurement of observer agreement for categorical data.
\newblock {\em Biometrics}, 33(1):159--174, 1977.

\bibitem{leclair2020improved}
Alexander LeClair, Sakib Haque, Lingfei Wu, and Collin McMillan.
\newblock Improved code summarization via a graph neural network.
\newblock In {\em Proceedings of the 28th International Conference on Program
  Comprehension}, pages 184--195, 2020.

\bibitem{lin2004rouge}
Chin-Yew Lin.
\newblock Rouge: A package for automatic evaluation of summaries.
\newblock In {\em Text summarization branches out}, pages 74--81, 2004.

\bibitem{liu2019automatic}
Zhongxin Liu, Xin Xia, Christoph Treude, David Lo, and Shanping Li.
\newblock Automatic generation of pull request descriptions.
\newblock In {\em 2019 34th IEEE/ACM International Conference on Automated
  Software Engineering (ASE)}, pages 176--188. IEEE, 2019.

\bibitem{luhn1958automatic}
Hans~Peter Luhn.
\newblock The automatic creation of literature abstracts.
\newblock {\em IBM Journal of research and development}, 2(2):159--165, 1958.

\bibitem{mani2001automatic}
Inderjeet Mani.
\newblock {\em Automatic summarization}, volume~3.
\newblock John Benjamins Publishing, 2001.

\bibitem{mihalcea2004textrank}
Rada Mihalcea and Paul Tarau.
\newblock Textrank: Bringing order into text.
\newblock In {\em Proceedings of the 2004 conference on empirical methods in
  natural language processing}, pages 404--411, 2004.

\bibitem{moreno2017automatic}
Laura Moreno and Andrian Marcus.
\newblock Automatic software summarization: the state of the art.
\newblock In {\em 2017 IEEE/ACM 39th International Conference on Software
  Engineering Companion (ICSE-C)}, pages 511--512. IEEE, 2017.

\bibitem{munaiah2017curating}
Nuthan Munaiah, Steven Kroh, Craig Cabrey, and Meiyappan Nagappan.
\newblock Curating github for engineered software projects.
\newblock {\em Empirical Software Engineering}, 22(6):3219--3253, 2017.

\bibitem{microsoft_gh}
Satya Nadella.
\newblock Microsoft + github = empowering developers.
\newblock
  \url{https://blogs.microsoft.com/blog/2018/06/04/microsoft-github-empowering-developers/},
  2018.

\bibitem{nassif2018automatically}
Mathieu Nassif, Christoph Treude, and Martin~P Robillard.
\newblock Automatically categorizing software technologies.
\newblock {\em IEEE Transactions on Software Engineering}, 46(1):20--32, 2018.

\bibitem{nenkova2012survey}
Ani Nenkova and Kathleen McKeown.
\newblock A survey of text summarization techniques.
\newblock In {\em Mining text data}, pages 43--76. Springer, 2012.

\bibitem{nenkova2005impact}
Ani Nenkova and Lucy Vanderwende.
\newblock The impact of frequency on summarization.
\newblock {\em Microsoft Research, Redmond, Washington, Tech. Rep.
  MSR-TR-2005}, 101, 2005.

\bibitem{nguyen_rocco_rubei_ruscio_2020}
Phuong~T. Nguyen, Juri~Di Rocco, Riccardo Rubei, and Davide~Di Ruscio.
\newblock An automated approach to assess the similarity of github
  repositories.
\newblock {\em Software Quality Journal}, 28(2):595–631, 2020.

\bibitem{page1963ordered}
Ellis~Batten Page.
\newblock Ordered hypotheses for multiple treatments: a significance test for
  linear ranks.
\newblock {\em Journal of the American Statistical Association},
  58(301):216--230, 1963.

\bibitem{pereira2015overview}
Dulce~G Pereira, Anabela Afonso, and F{\'a}tima~Melo Medeiros.
\newblock Overview of friedman’s test and post-hoc analysis.
\newblock {\em Communications in Statistics-Simulation and Computation},
  44(10):2636--2653, 2015.

\bibitem{prana2019categorizing}
Gede Artha~Azriadi Prana, Christoph Treude, Ferdian Thung, Thushari Atapattu,
  and David Lo.
\newblock Categorizing the content of github readme files.
\newblock {\em Empirical Software Engineering}, 24(3):1296--1327, 2019.

\bibitem{randolph2005free}
Justus~J Randolph.
\newblock Free-marginal multirater kappa (multirater k [free]): An alternative
  to fleiss' fixed-marginal multirater kappa.
\newblock {\em Online submission}, 2005.

\bibitem{rastkar2010summarizing}
Sarah Rastkar, Gail~C Murphy, and Gabriel Murray.
\newblock Summarizing software artifacts: a case study of bug reports.
\newblock In {\em 2010 ACM/IEEE 32nd International Conference on Software
  Engineering}, volume~1, pages 505--514. IEEE, 2010.

\bibitem{rennie2017self}
Steven~J Rennie, Etienne Marcheret, Youssef Mroueh, Jerret Ross, and Vaibhava
  Goel.
\newblock Self-critical sequence training for image captioning.
\newblock In {\em Proceedings of the IEEE Conference on Computer Vision and
  Pattern Recognition}, pages 7008--7024, 2017.

\bibitem{see2017_PointerGenerator}
Abigail See, Peter~J Liu, and Christopher~D Manning.
\newblock Get to the point: Summarization with pointer-generator networks.
\newblock In {\em Proceedings of the 55th Annual Meeting of the Association for
  Computational Linguistics (Volume 1: Long Papers)}, pages 1073--1083, 2017.

\bibitem{wan2018improving}
Yao Wan, Zhou Zhao, Min Yang, Guandong Xu, Haochao Ying, Jian Wu, and Philip~S
  Yu.
\newblock Improving automatic source code summarization via deep reinforcement
  learning.
\newblock In {\em Proceedings of the 33rd ACM/IEEE International Conference on
  Automated Software Engineering}, pages 397--407, 2018.

\bibitem{xiao2019extractive}
Wen Xiao and Giuseppe Carenini.
\newblock Extractive summarization of long documents by combining global and
  local context.
\newblock {\em arXiv preprint arXiv:1909.08089}, 2019.

\bibitem{zhang_2017}
Yun Zhang, David Lo, Pavneet~Singh Kochhar, Xin Xia, Quanlai Li, and Jianling
  Sun.
\newblock Detecting similar repositories on github.
\newblock In {\em 2017 IEEE 24th International Conference on Software Analysis,
  Evolution and Reengineering (SANER)}, pages 13--23, 2017.

\end{thebibliography}

\appendix

\section{LSP Test Coding Guidelines}
\begin{longtable}{p{.1\linewidth} p{.05\linewidth} p{.2\linewidth} p{.25\linewidth} p{.3\linewidth}}
    
    \textbf{Category} & \multicolumn {3}{p{.5\linewidth}}{\textbf{Rules}} & \textbf{Examples} \\
    \toprule
    \bottomrule
    \endfirsthead
    \textbf{Category} & \multicolumn {3}{p{.5\linewidth}}{\textbf{Rules}} & \textbf{Examples} \\
    \toprule
    \bottomrule
    \endhead
    
    \multirow{6}{*}{Language} & \textbf{L1} &  \multicolumn{2}{p{.45\linewidth}}{The coding language the repository is coded in.} & ``Phase unwrapping in native \textbf{Julia}" \\
            & \textbf{L2} & \multicolumn{2}{p{.45\linewidth}}{The coding language the repository's purpose is designed to be used for is present.} & ``Application tracing system for \textbf{Go}, based on Google's Dapper." \\
            & \textbf{L3} & \multicolumn{2}{p{.45\linewidth}}{A filename extension or equivalent, recognizable abbreviation for a coding language is present.} & ``\textbf{Py}torch implementation of convolutional neural network visualization techniques." \\
            & \textbf{L4} & \multicolumn{2}{p{.45\linewidth}}{The repository is a tutorial, course, or assignment in a specific coding language is mentioned in the description.} & ``Code samples for people who take part in my \textbf{Python} for Maya course." \\
            \cline{2-5}
            & \multicolumn{4}{p{.8\linewidth}}{A description does \textbf{not} contain a language if the only indication of the language is in the software technology \textbf{and} it does not contain a written indicator (i.e., L3).}\\
            & \multicolumn{4}{p{.8\linewidth}}{Note: there may be an overlap between Language and Purpose, in which case both are marked. I.e., ``Repository for \textbf{BASH} scripts and more."} \\
         \midrule
        \multirow{3}{.1\linewidth}{Software Technology} & \textbf{ST1} & \multicolumn{2}{p{.45\linewidth}}{There is at least one mention of a framework, library, tool, OS, protocol, or application.} & ``\textbf{.NET} cross-platform demo." \\
            & \textbf{ST2} & \multicolumn{2}{p{.45\linewidth}}{In cases where an ST is also an indicator for Language, the ST is also a Language and should be marked for both.} & ``Comprehensive implementations of Bootstrap 4 components and grid system for\textbf{ Vue.js}"\\
            \cline{2-5}
            & \multicolumn{4}{p{.8\linewidth}}{Note: a term is not a ST if it is a software concept (see Purpose category).}\\
         \midrule
         \multirow{5}{*}{Purpose} &  \textbf{P1} & \multicolumn{2}{p{.45\linewidth}}{ An implementation of key software concepts.} & ``Declarative routing for \textbf{React}." \\
            & \textbf{P2} & \multicolumn{2}{p{.45\linewidth}}{An implementation of a key software concept or a ST in a different language, framework, etc. (a ST might be directly mentioned).} & ``A TensorFlow implementation of Baidu's DeepSpeech architecture." \\
            & \textbf{P3} & \multicolumn{2}{p{.45\linewidth}}{An indication that the repository is used to hold code samples, code for tutorials/demos, or a website.} & ``This is the \textbf{code for \textbackslash{}DeepFakes\textbackslash{}} on Youtube"" \\
            & \textbf{P4} & \multicolumn{2}{p{.45\linewidth}}{What the repository does, what it is utilized for, or the added value to the end user.} & ``JavaScript program that will display information about the last 30 days' significant earthquakes." \\
            \cline{2-5}
            & \multicolumn{4}{p{.8\linewidth}}{Note: a description that only contains a ST alone is not a purpose. I.e., ``My Delphi project."} \\
        \toprule
        \bottomrule
    
\end{longtable}

\end{document}